\documentclass[seceq]{ptptex}

\usepackage{bm}

\title{
A New Analytical Method for Self-Force Regularization. I
}

\subtitle{Charged Scalar Particles in Schwarzschild Spacetime}

\author{Wataru \textsc{Hikida},$^1$
\footnote{E-mail: hikida@yukawa.kyoto-u.ac.jp}
Sanjay \textsc{Jhingan},$^2$
\footnote{E-mail: wtpsaxxj@lg.ehu.es}
Hiroyuki \textsc{Nakano},$^3$
\footnote{E-mail: denden@sci.osaka-cu.ac.jp}
Norichika \textsc{Sago},$^{4,5}$
\footnote{E-mail: sago@tap.scphys.kyoto-u.ac.jp}
Misao \textsc{Sasaki},$^1$
\footnote{E-mail: misao@yukawa.kyoto-u.ac.jp}
and Takahiro \textsc{Tanaka}$^4$
\footnote{E-mail: tama@scphys.kyoto-u.ac.jp} 
} 

\inst{
$^1$Yukawa Institute for
Theoretical Physics, Kyoto University,\\ Kyoto 606-8502, Japan
\\
$^2$Departamento de F\'{\i}sica Te\'orica, Universidad del
Pa\'{\i}s Vasco,\\ Apdo. 644, 48080, Bilbao, Spain
\\
$^3$Department of Mathematics and Physics, Graduate School of
Science,\\ Osaka City University, Osaka 558-8585, Japan
\\
$^4$Department of Physics, Graduate School of Science, Kyoto
University,\\ Kyoto 606-8502, Japan
\\
$^5$Department of Earth and Space Science, Graduate School of
Science,\\ Osaka University, Toyonaka 560-0043, Japan 
}

\abst{
We formulate a new analytical method for regularizing the
self-force acting on a particle of small mass $\mu$ orbiting a
black hole of mass $M$, where $\mu\ll M$. At first order in $\mu$,
the geometry is perturbed and the motion of the particle is
affected by its self-force. The self-force, however, diverges at
the location of the particle, and hence should be regularized. It
is known that a properly regularized self-force is given by the
tail part (or the $R$-part) of the self-field, obtained by
subtracting the direct part (or the $S$-part) from the full
self-field. The most successful method of regularization proposed
so far relies on the spherical harmonic decomposition of the
self-force, the so-called mode-sum regularization or mode
decomposition regularization. However, except for some special
orbits, no systematic analytical method for computing the
regularized self-force has been constructed. In this paper, utilizing a
new decomposition of the retarded Green function in the frequency
domain, we formulate a systematic method for the computation of
the self-force in the time domain. Our method relies on the
post-Newtonian (PN) expansion, but the order of the expansion can
be arbitrarily high. To demonstrate the essence of our method, in
this paper, we focus on a scalar charged particle on the
Schwarzschild background. Generalization to the gravitational
case is straightforward, except for some subtle issues related
with the choice of gauge (which exists irrespective of
regularization methods). 
}

\begin{document}

\maketitle

\section{Introduction}

We are now at the dawn of gravitational wave
astronomy/astrophysics. The interferometric gravitational wave
detectors LIGO~\cite{LIGO}, TAMA300~\cite{TAMA} and
GEO600~\cite{GEO} are currently in the early stage of their
operations, and VIRGO~\cite{VIRGO} is expected to be in operation
soon. Furthermore, R\&D studies of a space-based interferometer
project, LISA~\cite{LISA}, are in rapid progress. It is expected that
these
interferometers, after their final sensitivity goals are achieved,
will detect gravitational waves from compact star
binaries and compact stars orbiting super-massive black holes.

To fully utilize the information contained in the observed
gravitational wave data, and particularly for the purpose of precision
testing general relativity, it is essential to have accurate
theoretical predictions of the waveforms~\cite{Cutler}. For nearly
equal mass binaries, the standard PN approximation is a powerful
tool to compute the waveforms~\cite{Blanchet:2002av}. An
alternative method of computing the waveforms is the black hole
perturbation approach~\cite{RW,Zerilli,Chandra,Teukolsky,supple,
Sasaki-Tagoshi}. This approach is very effective, in particular,
when the mass ratio of the objects composing a binary is extreme.
In this paper, considering the case of such extreme mass
ratios, we propose a new method for calculating the corrections to
the force acting on the small mass body (which is treated as a
point particle) that are induced by the field generated by the
particle itself, the so-called self-force corrections.

In the black hole perturbation approach, one can appeal to the
energy-angular momentum balance argument to evaluate the radiation
reaction to the orbit of a particle, namely, by equating the rates
of change of the energy and angular momentum of the particle with
those carried away by the gravitational waves emitted by it.
However, the balance argument is not sufficient. First, the
radiation reaction to the Carter constant can not be calculated with 
this method. The Carter constant is the third constant of motion
of a test particle in the Kerr spacetime. The other two constants
of motion, the energy and angular momentum with respect to the
symmetry axis, are associated with the Killing vectors of the
background spacetime, and their rates of change may be evaluated
from the gravitational waves emitted to infinity or absorbed at the
black hole horizon. In contrast, there is no correspondence to any
such Killing vector in the case of the Carter constant. Hence, its
rate of change is not directly related to the waves emitted.
Second, and most importantly, the balance argument can yield only
time-averaged rates of change of the two constants of motion,
while there are many situations in which knowledge of the
actual radiation reaction force, as well as the so-called
conservative part of the self-force, at each instant of the orbital
motion becomes necessary.

Consider a particle having either a scalar, electromagnetic or
gravitational charge. The orbital motion of the particle will
create a field at first order in its charge, and the motion will
be affected by its self-field. This self-field, however, is 
divergent at the location of the particle. Hence, the force due to
this self-field is apparently ill-defined. It is known that the
self-force in the vicinity of the particle may be decomposed into
the so-called direct part and the tail part, and that the
correctly regularized self-force is given by the tail part. The
justification of this prescription is given in Ref.~\citen{DB} for the
scalar and electromagnetic cases, and in Ref.~\citen{MST} and ~\citen{QW}
for the gravitational case.

In the scalar case, the Klein-Gordon equation is hyperbolic from
the very beginning. In the electromagnetic and gravitational
cases, the field equations can be put into hyperbolic form by
choosing the Lorenz gauge (often called the `harmonic gauge'
in the gravitational case). In general, for a hyperbolic equation,
the retarded Green function $G^{\rm ret}(x,x')$ can be split into
two distinct pieces (at least locally when the two points, namely,
the field point $x$ and the source point $x'$ are sufficiently close),
which is called the Hadamard form. One piece has support
only on the future light-cone of $x'$, and the other piece has
support in the interior of the future light-cone of $x'$. The
former gives the direct part and the latter the tail part.
Recently, an equivalent but more elegant decomposition of the Green
function was proposed~\cite{DW03}, in which the direct part is
replaced by the $S$-part and the tail part by the $R$-part. The
$S$-part is defined by adding a piece that has support outside the
light-cone in such a way that it does not contribute to the
self-force when it is subtracted from the full field. The
remaining part is called the $R$-part, which now has support
outside the light-cone as well. The advantage of this new
decomposition is that the $S$-part is symmetric with respect to
$x$ and $x'$, and it satisfies the same equation as the retarded
Green function. This implies that the $R$-part now satisfies
a source-free, homogeneous equation.

Thus our task is to evaluate the tail or $R$-part of the field and
the self-force due to this. However, we do not have any systematic
method to compute the $R$-part directly. In contrast, there exist
several schemes to compute the full retarded field in
Schwarzschild and Kerr spacetimes~\cite{RW,Zerilli,Chandra,
Teukolsky,ManoTakasugi,ManoTak,supple,Sasaki-Tagoshi,Chrzanowski,
Ori:2002uv}. Therefore, what one can do is subtract the direct or
$S$-part from the full field to compute the regularized
self-force. There have been many investigation of this method
\cite{MNS,Barack:1999wf,NMS,BMNOS,Barack:2002bt,Barack:2002mh,Barack:2003mh}.

Because both the full field and the $S$-part diverge at the location
of the particle, it is necessary to develop a regularization
scheme to compute the difference between the two. The most
successful scheme of regularization that has emerged is the mode
decomposition (or mode-sum) regularization
\cite{MNS,Barack:1999wf}. The full field can be decomposed into
partial waves by using the spherical harmonics $Y_{\ell
m}(\Omega)$. The contribution to the force from each $\ell$-mode
does not diverge in the coincidence limit of the field point with
the location of the particle. Now, if we subtract the $S$-part
from the full field, before summing over $\ell$, the
divergence disappears. Hence, we can perform summation over $\ell$
to obtain the expression for the regularized force.

The $S$-part can be calculated only in the form of a local
expansion, that is, at field points sufficiently close to a point
on the orbit, and therefore it is necessary to extend it over a sphere
containing the orbital point to obtain the spherical harmonic
coefficients. In recent years, a method to carry out its harmonic
decomposition has been developed in the Schwarzschild
case~\cite{MNS,Barack:1999wf,NMS,BMNOS,Barack:2002bt,Barack:2002mh},
and it was recently extended to the case of the Kerr
background~\cite{Barack:2003mh}. The full field
is calculated using either the Regge-Wheeler-Zerilli or Teukolsky
formalism, which uses the spherical (or spheroidal in the Kerr
case) harmonic decomposition.

To this time, the regularized self-force has only been calculated numerically.
These numerical results are important. However, an analytical
understanding will be very useful. Analytical
results obtained to this time are restricted to the case of particles
orbiting in very special
orbits (mainly circular or radial infall), and there are no results
for the general case. This is primarily due to the mismatch in the
schemes used for evaluating the full Green function and the $S$-part. 
The Regge-Wheeler-Zerilli and Teukolsky formalism
rely heavily on the Fourier decomposition of the time-dependence by taking
full advantage of the stationarity of the background spacetime, and
therefore the full field is calculated in the frequency domain. 
Contrastingly, the $S$-part is in the time domain. If the orbit is specified
a priori, it is in principle possible to obtain the full field in
the time domain by explicitly performing the integration (or
summation) over the frequency $\omega$. However, in practice, the
explicit integration over $\omega$ is possible only in very
special cases, such as circular
orbits~\cite{Detweiler:2002gi,Burko:2000xx}.

In this paper, we propose a new method to go from the frequency domain
to time domain, and regularize analytically in order to calculate
the self-force completely analytically. Though we employ the PN
expansion method, it is formulated systematically in such a way
that the order of the expansion can be taken arbitrarily high, as
long as it is kept finite. In fact, our method is more effective
in the far zone, and thus it supplements the previous numerical
method, which is effective near the plunge. For simplicity, we restrict
our analysis to the case of a geodesic orbit in the background
spacetime, but removing this restriction is straightforward (although, 
of course, the equations become much more lengthy). Furthermore, we
focus on the case of a scalar charge in order to avoid the gauge
problem. There is a subtle problem associated with the choice of
gauge in the gravitational case~\cite{Sago:2002fe,NSS03},
 but our method for integration over $\omega$
is equally applicable to the gravitational case, despite this problem.

The key idea of the new approach developed in this paper is to
separate the retarded Green function in the frequency domain into
two distinct pieces, in analogy to the $S$-$R$ decomposition in
the space-time domain. We call them the $\tilde S$-part and the
$\tilde R$-part. The former contains all the singular terms to be
subtracted, while the latter satisfies the source-free,
homogeneous equation. In particular, once the PN order is
specified, the contributions from only a finite number of $\ell$ is
necessary to evaluate the $\tilde R$-part. The most important
point of this new decomposition is that the $\tilde S$-part in the
frequency domain is given in the form of a simple Taylor series
with respect to $\omega$ multiplied by $\exp[-i\omega(t-t')]$.
Therefore, the integration over $\omega$ can be performed easily
for such terms. They just produce $\delta(t-t')$ and its
derivatives. Using this technique, we can obtain the $\tilde
S$-part in the time domain relatively easily. Then, the
regularization is done by subtracting the $S$-part from the thus
obtained $\tilde S$-part in the time domain.

This paper is organized as follows. In \S\ref{sec:new-decomposition}, we
describe our new decomposition of
the retarded Green function in the frequency domain. In \S
\ref{sec:tilde-S}, focusing on the scalar case, we demonstrate our
regularization method. We first integrate the force due to
the $\tilde S$-part over $\omega$ to obtain the harmonic modes of
the force in the time domain, then subtract the $S$-part mode-by-mode,
and finally sum over $\ell$ to obtain the force due to the $(\tilde
S-S)$-part. We do not discuss the $\tilde R$-part, because it is
finite from the beginning. The final section, \S\ref{sec:conclusion}, is
devoted to conclusions and discussion. Some formulas and proofs of
several propositions used in the text are given in appendices. For
readers' convenience, formulas for 4PN order calculations can be found
at the {http://www2.yukawa.kyoto-u.ac.jp/\~{}misao/BHPC/}.

\section{New decomposition of the Green function in the frequency domain}
\label{sec:new-decomposition}

We consider a point scalar charge $q$ moving in the Schwarzschild
background
\begin{eqnarray}
ds^2 = - \left(1-{2M \over r}\right)dt^2 +\left(1-{2M \over
r}\right)^{-1}dr^2
 +r^2 \, (d\theta^2+\sin^2\theta d\varphi^2) \,,
\end{eqnarray}
where $\{x^{\alpha}\}=\{t, r, \theta, \phi\}$ are the
Schwarzschild coordinates, and $M$ is the black hole mass. The
full scalar field induced by this charged particle is given, using
the retarded Green function, by
\begin{eqnarray}
\psi^{\rm full}(x) &=& -q\int d\tau\, G^{\rm full}(x,z(\tau)) \,,
\label{eq:full-g}
\end{eqnarray}
where $\tau$ is the proper time of the particle, and $G^{\rm
full}(x,x')$ satisfies the Klein-Gordon equation,
\begin{eqnarray}
\nabla^\alpha \nabla_\alpha G^{\rm full}(x,x') &=&
-{\delta^{(4)}(x-x')\over \sqrt{-g}} \,, \label{eq:wave}
\end{eqnarray}
with retarded boundary conditions. The full Green function is
represented in terms of the Fourier-harmonic decomposition as
\begin{eqnarray}
G^{\rm full}(x,x') &=& \int {d\omega\over2\pi} \,
e^{-i\omega(t-t')} \sum_{\ell m} g^{\rm full}_{\ell m\omega}(r,r')
Y_{\ell m}(\theta,\phi)Y^*_{\ell m} (\theta',\phi')\,.
\label{eq:FHdeco-full-g}
\end{eqnarray}
Here, $Y_{\ell m}(\theta,\phi)$ are the ordinary spherical harmonics.
Then, Eq.~(\ref{eq:wave}) reduces to an ordinary
differential equation for the radial Green function as
\begin{eqnarray}
&&\left[\left(1-{2M\over r}\right){d^2\over dr^2} +{2(r-M)\over
r^2}{d\over dr}+\left({\omega^2\over\displaystyle
 1-{2M\over r}}-{\ell(\ell+1)\over r^2}\right)
\right]
g^{\rm full}_{\ell m\omega}(r,r')\cr&&
\hspace*{9.5cm}
 = -{1\over r^2}\delta(r-r')   \,. \label{eq:radial-eq}
\end{eqnarray}

The radial part of the full Green function can be expressed in
terms of homogeneous solutions of Eq.(\ref{eq:radial-eq}), which
can be obtained using a systematic analytic method developed in
Ref.~\citen{ManoTakasugi}{}. We have
\begin{eqnarray}
&&g_{\ell m\omega}^{\rm full}(r,r') = {-1\over W_{\ell
m\omega}(\phi_{\rm in}^{\nu},\phi_{\rm up}^{\nu})} \left(\phi_{\rm
in}^{\nu}(r)\phi_{\rm up}^{\nu}(r')\theta(r'-r) +\phi_{\rm
up}^{\nu}(r)\phi_{\rm in}^{\nu}(r')\theta(r-r')\right) \,,
\nonumber\\
&&W_{\ell m\omega}(\phi_{\rm in}^{\nu},\phi_{\rm up}^{\nu})
= r^2\left(1-{2M\over r}\right) \left[\biggl({d\over dr}\phi_{\rm
up}^{\nu}(r)\biggr)\phi_{\rm in}^{\nu}(r) -\biggl({d\over
dr}\phi_{\rm in }^{\nu}(r)\biggr)\phi_{\rm up}^{\nu}(r) \right].
\end{eqnarray}
Here, the in-going and up-going homogeneous solutions are
denoted, respectively, by $\phi_{\rm in}^{\nu}$ and $\phi_{\rm
up}^{\nu}$, and $\nu$ is called the `renormalized angular
momentum'~\cite{ManoTakasugi,ManoTak}, which is equal to $\ell$ in
the limit $M\omega \to0$.

We express the homogeneous solutions $\phi_{\rm in}^\nu$ and
$\phi_{\rm up}^\nu$ in terms of the Coulomb wave functions $\phi_{\rm
c}^{\nu}$ and $\phi_{\rm c}^{-\nu-1}$
~\cite{ManoTakasugi,Sasaki-Tagoshi};
\begin{eqnarray}
\phi_{\rm in}^{\nu} &=& \alpha_\nu\phi_{\rm
c}^{\nu}+\beta_{\nu}\,\phi_{\rm c}^{-\nu-1} \,,
\nonumber\\
\phi_{\rm up}^{\nu} &=& \gamma_{\nu}\,\phi_{\rm
c}^{\nu}+\delta_\nu\phi_{\rm c}^{-\nu-1} \,.
\end{eqnarray}
The properties and the relations of the coefficients
$\{\alpha_{\nu},\,\beta_{\nu},\,\gamma_{\nu},\,\delta_{\nu}\}$
are studied extensively in
Ref.~\citen{ManoTakasugi} and ~\citen{ManoTak}{}. (The function $\phi_{\rm
c}^\nu$ is denoted by $R_c^\nu$ in Ref.~\citen{ManoTakasugi}) Here, we
need to stress a remarkable property of the wave function
$\phi_c^{\nu}$, which becomes manifest when we consider the PN
expansion, i.e., when $\phi_c^\nu$ is expanded in terms of $z:=\omega r$
and $\epsilon:=2M\omega$, assuming they are small [$O(v)$ and $O(v^3)$,
respectively]. In the expression of its PN expansion,
$\Phi^\nu:=(2z)^{-\nu} \phi_c^{\nu}$ contains only terms that are
integer powers of $z$ and $\epsilon$, and there are no terms
like $\log z$. In fact, we find that this condition, that $\log z$
is absent, uniquely specifies a single solution of the radial
homogeneous equation. As is explained in Appendix
\ref{app:easy-method}, this fact can be used to compute
$\Phi^\nu$, simultaneously determining the eigenvalue $\nu$.
Furthermore, this PN expansion turns out to be a double Taylor
series expansion in $z^2$ and $\epsilon/z$, i.e., with only
positive powers of $\omega^2$. With the normalization
$\Phi^\nu\to1$, at the leading order, the expansion is
\begin{eqnarray}
\Phi^\nu & = &1-\frac{z^2}{2(2\ell+3)}-\frac{\ell \epsilon}{2z}
+\frac{z^4}{8(2\ell+3)(2\ell+5)}+ \frac{(\ell^2-5\ell-10)\epsilon
z}{4(2\ell+3)(\ell+1)}+\cdots,\cr \nu &=&
\ell-\frac{15\ell^2+15\ell-11}{2(2\ell-1)(2\ell+1)(2\ell+3)}\epsilon^2
+\cdots.\label{Phi}
\end{eqnarray}
The solution $\phi_{\rm c}^{-\nu-1}$ can be obtained through the
replacement $\ell \rightarrow -\ell-1$. The Wronskian of
$\phi^{\nu}_{\rm c}$ and $\phi^{-\nu-1}_{\rm c}$ becomes
\begin{eqnarray}
 &&\omega\,W_{\ell m\omega}(\phi_{\rm c}^{\nu},\phi_{\rm c}^{-\nu-1})
=-\frac{2\ell+1}{2}\cr&&
\quad\ +\frac{(496\ell^6+1488\ell^5+1336\ell^4
+192\ell^3-757\ell^2-605\ell+338)\epsilon^2}
{16(2\ell-1)^2(2\ell+1)(2\ell+3)^2}+\cdots.
 \end{eqnarray}
We note, however, that the general expression for the PN expansion
of $\phi_c^{-\nu-1}$ with this method (i.e., requiring the absence
of $\log z$ terms) becomes invalid at the $(\ell-1)$th PN
order. For small $\ell$ ($\leq\text{PN}+1$), the computation must
be done following the systematic method given in
Ref.~\citen{ManoTakasugi}{}. The respective results for the $\ell=0$
and $1$ cases are
\begin{eqnarray}
\Phi^\nu &=&
 \frac{7}{9}-\frac{7z^2}{54}-\frac{7\epsilon}{27z}+\frac{7z^4}{1080}
-\frac{14\epsilon z}{27}+\cdots,\cr 
\Phi^{-\nu-1} &=& -\frac{2z}{3\epsilon}j_0(z)+1+\frac{z^2}{18} +
\frac{\epsilon}{2z}+\cdots,\cr 
\nu &=& -\frac{7}{6}\epsilon^2+\cdots,\quad \omega
W=-\frac{49}{162}+\frac{23263}{29169}\epsilon^2+\cdots,
\end{eqnarray}
and
\begin{eqnarray}
\Phi^\nu &=& 1-\frac{z^2}{10}-\frac{\epsilon}{2z}+\frac{z^4}{280}
-\frac{7 \epsilon z}{20}+\cdots,\cr 
\Phi^{-\nu-1} &=&-\frac{30z^3}{19\epsilon} j_1(z) +1+\frac{29z^2}{38}
+\frac{\epsilon}{z}+\cdots, \cr 
\nu &=& 1-\frac{19}{30}\epsilon^2+\cdots,\quad 
\omega W = -\frac{3}{2}+\frac{117443}{216600}\epsilon^2+\cdots.
\end{eqnarray}
Here, $j_0(z)$ and  $j_1(z)$ are the spherical Bessel functions.

We now divide the Green function into two parts, as
\begin{eqnarray}\label{split}
&&g_{\ell m\omega}^{\rm full}(r,r')= g_{\ell m\omega}^{\tilde
S}(r,r') +g_{\ell m\omega}^{\tilde R}(r,r') \,,
\end{eqnarray}
where
\begin{eqnarray}
g_{\ell m\omega}^{\tilde S}(r,r')&=& {-1\over W_{\ell
m\omega}(\phi_{\rm c}^{\nu},\phi_{\rm c}^{-\nu-1})}\cr
&&
\times \biggl[
 \phi_{\rm c}^{\nu}(r)\phi_{\rm c}^{-\nu-1}(r')\theta(r'-r)
+\phi_{\rm c}^{-\nu-1}(r)\phi_{\rm c}^{\nu}(r')\theta(r-r') \biggr],
\label{eq:gC} \\
g_{\ell m\omega}^{\tilde R}(r,r')&=& {-1\over
(1-\tilde\beta_\nu\tilde\gamma_\nu) W_{\ell m\omega}
  (\phi_{\rm c}^{\nu},\phi_{\rm c}^{-\nu-1})}
\biggl[\tilde\beta_\nu\tilde\gamma_\nu \left(\phi_{\rm
c}^{\nu}(r)\phi_{\rm c}^{-\nu-1}(r') +\phi_{\rm
c}^{-\nu-1}(r)\phi_{\rm c}^{\nu}(r')\right)\cr&&
+\tilde\gamma_\nu \phi_{\rm c}^{\nu}(r)\phi_{\rm
c}^{\nu}(r') +\tilde\beta_\nu \phi_{\rm c}^{-\nu-1}(r)\phi_{\rm
c}^{-\nu-1}(r') \biggr]\,. \label{eq:gR}
\end{eqnarray}
Here, we have assumed that $\alpha_\nu\neq0$ and $\delta_\nu\neq0$,
and  we have also introduced the coefficients
$\{\tilde\beta_\nu,\,\tilde\gamma_\nu\}
:=\{\beta_\nu/\alpha_\nu,\,\gamma_\nu/\delta_\nu\}$. Using results
obtained in Ref.~\citen{ManoTakasugi}{}, we obtain the behavior of
the coefficients $\{\tilde\beta_\nu,\,\tilde\gamma_\nu\}$ in the PN
expansion as
\begin{eqnarray}
\tilde\beta_\nu=O(v^{6\ell+3})\,, \quad
\tilde\gamma_\nu=O(v^{-3})\,. \label{eq:betagamma}
\end{eqnarray}
The functions $\phi^\nu_c$ and $\phi^{-\nu-1}_c$ are,
respectively, of $O(v^\ell)$ and $O(v^{-\ell-1})$ (except at
$\ell=0$). Therefore, the three terms in the $\tilde R$-part of
the Green function become, respectively, of $O(v^{6\ell})$,
$O(v^{2\ell-2})$ and $O(v^{4\ell+2})$ relative to the $\tilde
S$-part.

The part that we need to consider in the regularization of the
force is just $\tilde S$. There is no divergence associated with
the remaining $\tilde R$-part, and it terminates at finite $\ell$
as long as we restrict our consideration to finite PN order. Moreover,
the $\tilde R$-part satisfies the homogeneous radial equation. This fact will
be an additional advantage of the present method when we consider
extension to the case of gravity. Because the $\tilde R$-part is a
homogeneous solution, we can apply Chrzanowski's
method~\cite{Chrzanowski} to reconstruct the metric perturbations
even in the frequency domain. We discuss this point in more
detail in a separate paper.

\section{Computation of the $\tilde S$-part}\label{sec:tilde-S}

We now compute the force due to the $\tilde S$-part for a general
orbit. When expanded in terms of the spherical harmonics, the
force corresponding to the $S$-part (the $S$-force) is known to
take the form~\cite{Barack:1999wf}
\begin{equation}
 \lim_{x\to z_0} F^{S}_{\alpha,\ell}
  =A_\alpha L+B_\alpha + D_{\alpha,\ell},
\end{equation}
where $F^{S}_{\alpha \ell}$ is the $\ell$-mode of the $S$-force,
$L=\ell +1/2$, and $A_{\alpha}$ and $B_\alpha$ are independent of
$L$. When summed over $\ell$, the $A$-term gives rise to a
quadratic divergence, and the $B$-term diverges linearly. For large
$\ell$, $D_{\alpha \ell}$ is at most of $O(L^{-2})$, and hence it 
contains no divergence. Because the $S$-part can be calculated only
locally, its extension to the entire sphere involves some ambiguity. As
a result, the coefficient of each $\ell$-mode, $D_{\alpha \ell}$,
depends on the method of extension, but the final result after
summation over $\ell$, which is determined by the local behavior
of the field near the source location, does not. It is known that
\begin{equation}
\sum_{\ell=0}^{\infty} D_{\alpha,\ell} =0.
\end{equation}
The difference between the $S$-force and the $\tilde S$-force
should be finite, because the $\tilde R$-force is finite. Thus, in
general, the $\tilde S$-force must take the same form as the
$S$-force
\begin{equation}
 \lim_{x\to z_0} F^{\tilde S}_{\alpha,\ell}
  =A_\alpha L+B_\alpha + \tilde D_{\alpha,\ell}\,.
\label{standardform}
\end{equation}
Below we confirm explicitly that both $A_{\alpha}$ and $B_{\alpha}$ for
the $\tilde S$-force coincide with those for the $S$-force. Therefore,
the force due to the $\tilde S$-part minus the $S$-part, which is
finite, is given by
\begin{equation}
F^{\tilde S-S}_{\alpha}= \sum_{\ell=0}^{\infty} \lim_{x\to z_0}
\left(F^{\tilde S}_{\alpha,\ell}-F^{S}_{\alpha,\ell}\right)
   = \sum_{\ell=0}^{\infty} \tilde D_{\alpha,\ell}.
\end{equation}

\subsection{Force in the time domain}
\label{sub:force-TD}

To obtain an expression for the $\tilde S$-force in the form
of Eq.~(\ref{standardform}), it is necessary to perform the $\omega$
integration explicitly. Here, the key fact is that there appears no
fractional power of $\omega$ in the $\tilde S$-part. This is
because we have chosen $\phi_c^\nu$ and $\phi_c^{-\nu-1}$ as the
two independent basis functions. As noted above, except for the
overall fractional powers $z^\nu$ and $z^{-\nu-1}$, they contain
only the terms with positive integer powers of $\omega^2$. When we
consider a product of these two functions, $\omega$ contained in
the overall factors $z^\nu$ and $z^{-\nu-1}$ just produces
$\omega^{-1}$, which is canceled by $\omega$ from the inverse of
the Wronskian. Thus, $g^{\tilde S}_{\ell m\omega}(r,r')$ is
expanded as
\begin{equation}
 g^{\tilde S}_{\ell m\omega}(r,r')
   =\sum_{k=0}^{\infty} \omega^{2k} {\cal G}_{\ell m k}(r,r').
\end{equation}
Therefore, the integration over $\omega$ can be performed easily,
as we shall show now explicitly. The Fourier transform of
$\omega^{2n}$ simply produces
\[
 \int d\omega\, \omega^{2n} e^{-i\omega (t-t')}=
    2\pi{ (-1)^n \partial_{t'}^{2n} }\delta (t-t').
\]
Differentiation of the delta function in the expression above can
be integrated by parts to act on the source term. Thus, we can
express the $\tilde S$-force in the time domain as
\begin{equation}
 F_{\alpha,\ell}^{\tilde S}= q^2
P_\alpha{}^\beta
\lim_{x\to z(t)}
\nabla_\beta \sum_{m, k}
  (-1)^k(\partial_t)^{2k} {d\tau(t)\over dt}
     {\cal G}_{\ell m k}(r,z^r(t)) Y_{\ell m}(\theta, \varphi)
     Y_{\ell m}^* (z^\theta (t), z^\varphi(t)).
\label{forceS-S}
\end{equation}
Here we have inserted the projection tensor
$P_\alpha{}^\beta=\delta_\alpha{}^\beta+u_\alpha u^\beta$, where
$u^\alpha$ is the four-velocity, so that
the normalization $u^\alpha u_\alpha=-1$ is maintained.
Now that we have the $\tilde S$-force given in the time domain,
the meaning of the coincidence limit $r\to z^r(t)$ is transparent.

Here one can set $z^\theta (t)\equiv \pi/2$ without loss of
generality. Then, nothing that $\partial_\varphi Y_{\ell m} (\pi/2,
\varphi)=i m Y_{\ell m}(\pi/2, \varphi)$ and $\partial_\varphi
Y_{\ell m}^*(\pi/2,\varphi)=-imY_{\ell m}^*(\pi/2,\varphi)$, the
summation over $m$ can be done by using the formulas
\begin{eqnarray}
&& \sum_{m=-\ell}^\ell m^{j}\, \left\vert Y_{\ell m}
   \left({\pi\over 2},\varphi \right)\right\vert^2
=\left\{
\begin{array}{ll}
  \displaystyle \lambda_{(j/2)}(\ell),~
   & $for $ j=$even$, \cr
  ~0, & $for $ j=$odd$,
\end{array}\right. \cr
&& \sum_{m=-\ell}^\ell m^{j} \,\partial_\theta Y_{\ell m}
   \left(\theta,\varphi \right)\Bigr|_{\theta=\pi/2}\,
Y_{\ell m}^*\left({\pi\over 2},\varphi \right) =0,
\end{eqnarray}
where $\lambda_{(n)}(\ell)$ is a polynomial function of $\ell$ of
order $n+1$ defined by
\begin{equation}
\sum_{n=0}^\infty \frac{\lambda_n z^{2n}}{n!}=
    {2\ell+1\over 4\pi} e^{\ell z}\,_2F_1\left(
    {1\over 2}, -\ell ;1 ;1-e^{-2z}\right).
\end{equation}

\subsection{Separation of the $A$-term}
\label{sub:separation-A}

Before performing the operation discussed in the preceding
subsection, we separate the $A$-term from the other contributions.
This can be done easily by using the fact that only the
$A$-term has a jump in its values at the coincidence limit,
depending on the direction from which the source point is approached. We
divide the $\tilde S$-part Green function into 
symmetric and anti-symmetric parts as
\begin{equation}
  g^{\tilde S}_{\ell m\omega}(r,r')
   =   g^{\tilde S(+)}_{\ell m\omega}(r,r')
    +{\rm sgn}(r-r')
       g^{\tilde S(-)}_{\ell m\omega}(r,r'),
\end{equation}
where ${\rm sgn}(y)=\pm 1$ for $y\gtrless0$, and
\begin{eqnarray}
g_{\ell m\omega}^{\tilde S(\pm)}(r,r')  & = & {-1\over 2W_{\ell
m\omega}(\phi_{\rm c}^{\nu},\phi_{\rm c}^{-\nu-1})} \left[
\phi_{\rm c}^{\nu}(r)\phi_{\rm c}^{-\nu-1}(r') \pm \phi_{\rm
c}^{-\nu-1}(r)\phi_{\rm c}^{\nu}(r')\right]. \label{eq:symgC}
\end{eqnarray}
Then the force due to the anti-symmetric part must coincide
exactly with the $A$-term. We know that the $A$-term for the
$S$-part has a simple form proportional to $L$, while the
expression for the $\tilde S$-force given in terms of the Green
function is more complicated. The reason why such a simple
result for the $A$-term is recovered is explained in Appendix
\ref{app:A-term}. We concentrate now on the symmetric part, which
is responsible for the $B$ and $D$-terms.

\subsection{The $(\tilde{S}-S)$-part of the force}
\label{sub:S-tildeS}

The result for the $\tilde{S}(+)$-part of the force is
\begin{eqnarray}
 && F_{t,\ell}^{\tilde{S}(+)}=\frac{q^2 u^r}{4\pi r_0^2}
\sum_{n=0}^{\infty}K_{t}^{(n)}(\ell),\quad
 F_{\theta,\ell}^{\tilde{S}(+)}=0,\quad
 F_{\varphi,\ell}^{\tilde{S}(+)}=\frac{q^2 u^r\mathcal{L}}{4\pi r_0^2}
 \sum_{n=0}^{\infty}K_\varphi^{(n)}(\ell),
\end{eqnarray}
and
\begin{equation}
F_{r,\ell}^{\tilde{S}(+)}=-\frac{\mathcal{E}}{u^r(1-2M/r_0)}
F_{t,\ell}^{\tilde{S}(+)}-\frac{\mathcal{L}}{u^r r_0^2}
F_{\varphi,\ell}^{\tilde{S}(+)},
\end{equation}
where the coefficients $K_{\alpha,\ell}^{(i)}$, whose upper index
$(i)$ represents the PN order, are formally given by
\begin{eqnarray}
&& K_{t,\ell}^{(n)}=\sum_{i+j+k=n} d_{t}^{\ (ijk)}(\ell)\
(\delta_{\mathcal{E}})^i
\left(\frac{\mathcal{L}^{2}}{r_0^2}\right)^jU^k,\cr
&& K_{\varphi,\ell}^{(n)}=\sum_{i+j+k=n}
d_{\varphi}^{\ (ijk)}(\ell)\  (\delta_{\mathcal{E}})^i
\left(\frac{\mathcal{L}^{2}}{r_0^2}\right)^jU^k.
\end{eqnarray}
Here, the quantities $d_\alpha$ are some functions of $\ell$, $r_0\equiv z^r(t_0)$,
$\delta_{\mathcal{E}}\equiv1-\displaystyle\frac{1}{\mathcal{E}^2}$,
$U\equiv\displaystyle\frac{M}{r_0},$ and $\mathcal{E}$ and
$\mathcal{L}$ are, respectively, the energy and the angular
momentum of the particle. To obtain these expressions, we have
used the first integrals of the geodesic equations,
\begin{eqnarray}
 \left(\frac{dz^r(t)}{dt}\right)^2 &=& \left(1-\frac{2M}{z^r(t)}\right)^2
-\frac{1}{\mathcal{E}^2}\left(1-\frac{2M}{z^r(t)}\right)^3
\left(1+\frac{\mathcal{L}^2}{z^r(t)^2}\right),\cr
\frac{d{z^\varphi(t)}}{dt} &=&
\frac{\mathcal{L}}{\mathcal{E}}\frac{1}{z^r(t)^2}
\left(1-\frac{2M}{z^r(t)}\right),\quad \frac{dt}{d\tau} =
\frac{\mathcal{E}}{1-2M/z^r(t)}\label{geodesic},
\end{eqnarray}
and we have also reduced higher-order derivatives with respect to $t$ by using the
equations of motion
\begin{eqnarray}
 \frac{d^2z^r(t)}{dt^2} &=&
\frac{2M}{z^r(t)^2}\left(1-\frac{2M}{z^r(t)}\right)
+\frac{\mathcal{L}^2}{\mathcal{E}^2z^r(t)^3}
\left(1-\frac{2M}{z^r(t)}\right)^3
\cr && \qquad
-\frac{3M}{\mathcal{E}^2z^r(t)^2}
\left(1-\frac{2M}{z^r(t)}\right)^2\left(1+\frac{\mathcal{L}^2}{z^r(t)^2}\right)
.\label{geoeq}
\end{eqnarray}
As long as we ignore corrections that are higher order in $\mu$,
we can assume that the orbit is momentarily geodesic. Here, we find
that the $\tilde S$-part force is written solely in terms of the
orbit at the location of the particle (that is, no tails!).
Hence, for the correction at lowest order in $\mu$, the force
coming from the $(\tilde S-S)$-part is also written in terms of the
position and the velocity of the particle. We do not have any
terms with a positive power of $\ell$, as expected. However,
this cancellation looks rather miraculous in the present
formulation. In Appendix \ref{app:large-ell}, we give a brief
explanation of why the terms with positive powers of $\ell$ are absent.
From the asymptotic behavior for large $\ell$, we can read off the
coefficients $B_{\alpha}$. These coefficients are identical to the
results obtained previously.~\cite{BMNOS,MNS}
As mentioned earlier, a separate treatment is necessary for small
$\ell$. To summarize, the $(\tilde{S}-S)$-part of the force is given by
\begin{eqnarray}
 F^{\tilde{S}-S}_\alpha &=& \sum_{\ell=0}^{\infty}\lim_{x\to z_0}
\left(F^{\tilde{S}}_{\alpha,\ell}-F^{S}_{\alpha,\ell}\right)
=  \sum_{\ell=0}^{\infty}
\left(F^{\tilde{S}}_{\alpha,\ell}-
A_{\alpha}\left(\ell+\frac{1}{2}\right)-B_{\alpha}\right)\cr
&=& \sum_{\ell=0}^{\infty}\left(F^{\tilde{S}(+)}_{\alpha,\ell}
-B_{\alpha}\right)
\end{eqnarray}

The summation over $\ell$ is performed by using the decomposition into
partial fractions and the formulas
\begin{eqnarray}
 \sum_{\ell=1}^\infty {1\over \ell^{2n}} & = &
  2^{2n-1}\pi^{2n} {B_n\over (2n)!},\quad
 \sum_{\ell=1}^\infty  {1\over (\ell-1/2)^{2n}}
   =  2^{2n-1}(2^{2n}-1)
\pi^{2n} {B_n\over (2n)!},\quad
\end{eqnarray}
where $n$ is a positive integer, and $B_n$ is the Bernoulli number
defined by
\begin{equation}
 {x\over e^x -1}+{x\over 2} -1=\sum_{n=1}^\infty 2^{2n}(2^{2n}-1)
        {B_n\over (2n)!} x^{2n}.
\end{equation}
The reason why odd powers of $\ell$ do not arise is as follows. If
we have a term like
\[
\sum_{n=0}^\infty {1\over (\ell+k/2)^n} ,
\]
then, due to the symmetry under $\ell\to -(\ell+1)$, we also have a term
\[
\sum_{n=0}^\infty {1\over (-\ell-1+k/2)^n} ,
\]
where $k$ is an integer. If $n$ is odd, these two contributions
combine to leave a summation of finite terms. Hence, there
remains no infinite summation of odd power terms in the final
expression.

The result for the $(\tilde{S}-S)$-part of the scalar self-force is
\begin{eqnarray}
 && F_t^{\tilde{S}-S}=\frac{q^2 u^r}{4\pi r_0^2}
\sum_{n=0}^{\infty}C_{t}^{\tilde{S}-S(n)},\quad
 F_\theta^{\tilde{S}-S}=0,\quad
 F_\varphi^{\tilde{S}-S}=\frac{q^2 u^r\mathcal{L}}{4\pi r_0^2}
 \sum_{n=0}^{\infty}C_\varphi^{\tilde{S}-S(n)},\quad
\end{eqnarray}
and
\begin{equation}
F_r^{\tilde{S}-S}=-\frac{\mathcal{E}}{u^r(1-2M/r_0)}F_t^{\tilde{S}-S}
-\frac{\mathcal{L}}{u^r r_0^2}F_\varphi^{\tilde{S}-S},
\end{equation}
where the coefficients $C_\alpha^{\tilde{S}-S(i)}$, whose upper index,
$i$, represents the PN order, are formally given by
\begin{eqnarray}
&& C_t^{\tilde{S}-S(n)}=\sum_{i+j+k=n} e_t^{(ijk)} (\delta_{\mathcal{E}})^i
\left(\frac{\mathcal{L}^{2}}{r_0^2}\right)^jU^k,\cr
&& C_\varphi^{\tilde{S}-S(n)}=\sum_{i+j+k=n} e_\varphi^{(ijk)} (\delta_{\mathcal{E}})^i
\left(\frac{\mathcal{L}^{2}}{r_0^2}\right)^jU^k.
\end{eqnarray}
Here $e_\alpha$ are some constant numbers.

Once we obtain the general expression for the $(\tilde S-S)$-part of
the force, computation of the remaining $\tilde R$-part is
rather easy, because only terms up to a finite value of $\ell$
contribute to the force for a given PN order.
Then, the $R$-part of the force, which is what we want in the end, is
given by
\begin{eqnarray}
 F^{R}_\alpha = F^{\tilde{S}-S}_\alpha + F^{\tilde{R}}_\alpha.
\end{eqnarray}
\vspace*{5mm}

Now we discuss a technical but important property of the $\tilde
S$-part of the Green function. The two independent radial
functions at leading order in $\epsilon$ are given by the
spherical Bessel functions $j_{\ell}(z)$ and $n_{\ell}(z)$. Both
$\phi_c^\nu$ and $\phi_c^{-\nu-1}$ are given by linear
combinations of these two independent solutions. Up to
$O(\epsilon^0)$, we have $\phi_c^\nu\propto j_{\ell}(z)$, while
$\phi_c^{-\nu-1}\propto \epsilon^{-1} j_{\ell}(z)+C_\ell\,n_\ell(z)$,
where $C_\ell$ is a constant of order unity. Here, in
passing, we note that the leading term of $\phi_c^{-\nu-1}$ in the
PN expansion comes not from the term $\epsilon^{-1}j_\ell(z)$ but
from $n_{\ell}(z)$ for $\ell\geq2$. This is because the ratio of
the two terms is $j_{\ell}(z)/(\epsilon n_{\ell}(z))\propto
z^{2\ell-2}=O(v^{2\ell-2})$ in the PN expansion. At first glance,
the existence of the term $\epsilon^{-1}j_\ell(z)$ seems
problematic, since it would naively lead to a term in the Green
function that behaves as $1/M$. Collecting the leading terms in
$\epsilon$, we find that the Green function has a term
proportional to $1/M$ of the form
\begin{equation}
 \propto {1\over M}\sum_{\ell m}\int {d\omega} e^{-i\omega(t-t')}
   j_{\ell}(\omega r) j_{\ell}(\omega r')
   Y_{\ell m} (\Omega) Y^*_{\ell m} (\Omega'). \nonumber
\end{equation}
This expression is identical  the radiative Green function in
Minkowski space, except for an additional multiplicative factor
$1/(M\omega)$. Therefore, after summation over $\ell$ and $m$ and
integration over $\omega$, we find that this part is
\begin{eqnarray*}
\propto&& \int dt' \left[{\delta(t-t'-|{\bf x}-{\bf x}'|)\over
                     |{\bf x}-{\bf x}'|}
                -{\delta(t-t'+|{\bf x}-{\bf x}'|)\over
                     |{\bf x}-{\bf x}'|}\right]\cr
&&\qquad  ={\theta (t-t'+|{\bf x}-{\bf x}'|)
    -\theta (t-t'-|{\bf x}-{\bf x}'|)
            \over |{\bf x}-{\bf x}'|}.
\end{eqnarray*}
At lowest order in $M$, the trajectory of a particle is a
straight line in the Minkowski background. Because the above
expression for the leading part of the Green function is Lorentz
invariant, we can choose this straight line to be static,
without loss of generality. Then, it is easy to see that the
component of the field proportional to $1/M$ is
constant. Hence, this part does not contribute to the force.

Another important property of the force due to the
$(\tilde S-S)$-part is that it contains only the conservative part of the
force. To show this, we use the fact that the equations of motion
take the form $\dot u^r=$ (a function even in $u^r$) and
 $\dot {\cal E}= \dot {\cal L}=0,$ at leading order in $\mu$
[see Eqs.~(\ref{geodesic}) and (\ref{geoeq})].
Here, the dot represents differentiation with respect to $t$.
Recalling the general expression for the $\tilde S$-force in
Eq.~(\ref{forceS-S}), we see that the $t$-component contains an
odd number of time derivatives, $\partial^{2k+1}/\partial
t^{2k+1}$, the $r$-component an even number of time derivatives
plus one radial derivative, $\partial^{2k+1}/(\partial
t^{2k}\partial r)$, and the $\varphi$-component an even number of
time derivatives plus one $\varphi$ derivative,
$\partial^{2k+1}/(\partial t^{2k}\partial\varphi)$. Now, using the
equations of motion, the $t$ derivative can be replaced by
\begin{eqnarray*}
\frac{\partial}{\partial t} &=&\dot z^r(t)\frac{\partial}{\partial
z^r} +\ddot z^r(t)\frac{\partial}{\partial \dot z^r} +\dot
z^{\varphi}(t)\frac{\partial}{\partial z^\varphi}
\\
&=& \dot z^r(t)\frac{\partial}{\partial z^r} +\ddot
z^r(t)\frac{\partial}{\partial \dot z^r} -im\,\dot z^\varphi(t)\,,
\end{eqnarray*}
and the $\varphi$ derivative by
\begin{eqnarray*}
\frac{\partial}{\partial\varphi}=+im\,.
\end{eqnarray*}
Notice that one differentiation with respect to $t$ or $\varphi$
changes the total power of $u^r$ and $m$ by an odd number, while a
differentiation with respect to $r$ does not. Notice also that
only the terms even in $m$ remain after summation over $m$.
Therefore the $\ell$-mode of the $\tilde S$-force takes the form
\begin{equation}
F^{\tilde S}_{t\,\ell}={\cal F}_{t\,\ell}(r,\mathcal{E},\mathcal{L})\,u^r,
\quad F^{\tilde S}_{r\,\ell}
={\cal F}_{r\,\ell}(r,\mathcal{E},\mathcal{L}),
 \quad F^{\tilde S}_{\varphi\,\ell}
={\cal F}_{\varphi\,\ell}(r,\mathcal{E},\mathcal{L})\,u^r\,.
\end{equation}
The $S$-part of the force is known to have exactly the same form.
This implies that the $(\tilde S-S)$-part of the force also takes
the same form. Thus, after summing over $\ell$, we conclude that
the final form of the $(\tilde S-S)$-part of the force is
\begin{eqnarray}
F^{\tilde S-S}_t={\cal F}_t (r,\mathcal{E},\mathcal{L})\,u^r,\quad
F^{\tilde S-S}_r={\cal F}_r (r,\mathcal{E},\mathcal{L}), \quad
F^{\tilde S-S}_\varphi
={\cal F}_\varphi(r,\mathcal{E},\mathcal{L})\, u^r\,.
\label{tilS-Sform}
\end{eqnarray}

We can now explicitly show that the above form of the force
implies the absence of a dissipative reaction effect. In other words,
the force is conservative. The
equations of motion to $O(\mu^2)$ are given by
\begin{equation}
 \mu{D\over d\tau} \tilde u^\mu=F^\mu,
\end{equation}
where $\tilde u^\mu$ is the perturbed four velocity and $D/d\tau$
is the covariant derivative. Then, we obtain the evolution equation for
the perturbed energy $\tilde{\cal E}:=-\mu\,\hat t_{\mu} \tilde u^\mu$ as
\begin{equation}
 {d\tilde{\cal E}\over d\tau}
 =-\mu{D\over d\tau}(\hat t_\mu \tilde u^\mu)
     =-\hat t_\mu F^\mu = -{\cal F}_t (r) {dr\over d\tau},
\end{equation}
where $\hat t^\mu=(\partial_t)^\mu$
is the time-like Killing vector. This equation
is integrated to give
\begin{equation}
  \tilde {\cal E}={\cal E}-\int^r \, {\cal F}_t(r) dr.
\end{equation}
Here, ${\cal E}$ is an integration constant, which we can interpret
as the unperturbed energy.
In the same manner, for the perturbed angular momentum ${\cal L}$, we obtain
\begin{equation}
  \tilde {\cal L}={\cal L}+\int^r \, {\cal F}_\varphi (r) dr.
\end{equation}
Thus we find that there is no cumulative effect on the
evolution of the energy and angular momentum of the particle.
In other words,
a force of the form~(\ref{tilS-Sform}) preserves the presence of the
constants of motion ${\cal E}$ and ${\cal L}$.
Concerning the radial motion, $\tilde{u}^r$ can be expressed
in terms of $\mu\,\tilde{u}_t=\tilde{\cal E}$ and
$\mu\,\tilde{u}_\varphi=\tilde{\cal L}$
by using the normalization condition of the
four velocity. We have 
\begin{eqnarray}
\mu\,\tilde u^r=\pm\left[
\tilde{\cal E}^2-(1-2M/r)\left(1+\tilde{\cal L}^2/r^2\right)
\right]^{1/2}.
\end{eqnarray}
Thus, $\tilde u^r$ is obtained as a function of $r$.

\section{Conclusions and discussion}\label{sec:conclusion}

The present work is an attempt to make progress toward a more \emph{realistic}
(calculation effective) analytic scheme for constructing orbits,
taking into account radiation reaction effects. The key idea
proposed in this paper is a new decomposition of the Green
function into $\tilde S$ and $\tilde R$-parts,
given in Eq.~(\ref{split}). This new decomposition relies on a systematic
analytic approach to the black hole perturbation developed in
Refs.~\citen{Sasaki-Tagoshi,ManoTakasugi} and ~\citen{ManoTak}. The new
decomposition is not identical to the usual $S$ and $R$
decomposition~\cite{DW03}, but they share certain properties. The
$\tilde S$-part is singular and symmetric, and it satisfies the
same inhomogeneous equation as the Green function. The $\tilde
R$-part is regular, and it satisfies the source-free equation.
Considering a scalar charged particle, we showed that the $\tilde
S$-part of the self-force can be evaluated analytically in the
time domain and that it yields the same regularization parameters
$A_{\alpha}$ and $B_{\alpha}$ in the mode-decomposition
regularization as the usual $S$-part. This implies that the $\tilde
S$-part contains all the singular behavior of the original
$S$-part. Also, we showed that the self-force due to the $(\tilde
S-S)$-part is conservative. Moreover, we found that the $\tilde
R$-part of the force valid up to the $(\ell+0.5)$th Post-Newtonian
(PN) order can be obtained by taking account of only the spherical
harmonic modes up to $\ell$-th order.

The analysis here is restricted to the self-force due to a scalar
field for its simplicity. The extension to the electromagnetic
case is straightforward, although the computation of the force
from the master variable of the perturbations~\cite{Teukolsky} becomes
more tedious.
The gravitational case, however, is more complicated, because in that
case, the self-force is gauge dependent. In the scalar case, as is manifest
from our calculation, the computation of the $S$-part is in fact
unnecessary if we make use of the fact that the non-singular
part of the regularization parameter $D_{\alpha \ell}$ for the
$S$-part vanishes after summing over $\ell$ modes. By virtue of
this fact, the simple prescription of subtracting $A$- and $B$-terms
is valid in this case. In the case of gravity, before subtracting
the $S$-part from the $\tilde S$-part of the force, we need to
adjust the gauges that are originally different. A natural
prescription for this is to transform the $S$-part, originally
given in Lorenz gauge, to the gauge in which the $\tilde S$-part
is computed. Due to this gauge transformation, the parameter
$D_{\alpha \ell}$ as well as the other regularization parameters
will be altered. Then the contribution to the self-force from
$D_{\alpha \ell}$ will not vanish in general. This needs to be
studied in more detail.

Related to the gauge dependence of the gravitational self-force,
there arises a conceptual question: What kind of gauge-independent
concepts are contained in this otherwise gauge-dependent
quantity? There are several constants of motion for geodesics in
the background black hole spacetime. These constants of motion
evolve after we incorporate the self-force. However, the secular
change of ``the constants of motion'' has a gauge-invariant
meaning. It can be evaluated by merely taking account of the force due to
the $\tilde R$-part, or we can simply use the radiative Green
function~\cite{Mino2003}. In addition to the information regarding the
constants of motion, the self-force may contain
other gauge-invariant information.
This question needs to be answered, although it is not a problem
specific to our present method. We may find that the physical
information contained in the gauge-dependent self-force by itself
is very limited. Even if this is the case, calculation of the self-force
will be a necessary step to develop a black hole perturbation theory
to second order in the mass of the orbiting particle.

One of the main advantages of our method is that it allows a  successful
implementation of a systematic post-Newtonian expansion technique
in the black hole perturbation.
There could be criticism of our approach in regard to the
limitation of the PN expansion itself.
The black hole perturbation is considered to be a method
complimentary to the standard post-Newtonian approximation.
In this sense, one might think that there is no point in
using the PN expansion in the black hole perturbation approach.
However, we should stress that the PN expansion
of the perturbation in the black hole spacetime can be systematically
extended to an arbitrarily high order without any conceptual
difficulties. Hence, if we can make use of this advantage,
problems far beyond the validity of the finite
order standard post-Newtonian approximation can be investigated.

In an actual computation, the achievable PN order will be limited.
Then, the question will be the speed of
the convergence of the PN expansion. In this regard, there are a couple of
encouraging pieces of evidence. First, we mention the issue of the
innermost stable circular orbit (ISCO).
At present, the calculation of the orbital frequency at ISCO up to 3PN order
has been made~\cite{Jaranowski}. The result is in rather
good agreement with the numerical result for an equal-mass
binary~\cite{Gourgoulhon}. Second, we mention the energy loss
rate of a particle in a circular orbit in the Schwarzschild black
hole evaluated from the asymptotic waveform~\cite{supple}.
 Although the convergence of the PN expansion becomes slower and slower for
a smaller orbital radius, there is no evidence of the failure 
converge, even at ISCO~\cite{Damour}.
Hence, if we can develop a systematic
method of evaluating the gravitational self-force in the PN expansion,
its range of validity should be very wide.

\section*{Acknowledgements}

We would like to thank Y. Mino and T. Nakamura for useful
discussions. We also thank all the participants at the 6th Capra
meeting and the Post Capra meeting at the Yukawa Institute,
 Kyoto University (YITP-W-03-02).
HN is supported by a JSPS Research Fellowship for Young Scientists
(No.~5919). SJ acknowledges support under a Basque government postdoctoral
fellowship. This work was supported in part by Monbukagaku-sho
Grants-in-Aid for Scientific Research (Nos.~14047212, 14047214 and
12640269), and by the Center for Gravitational Wave Physics, PSU,
which is funded by NSF under the Cooperative Agreement PHY~0114375.

\appendix
\section{An Easy Way to Find $\phi^\nu$ for Sufficiently Large
$\ell$}\label{app:easy-method}

Consider a radial function $\phi^\nu(z)$ which satisfies
\begin{eqnarray}
\left[\left(1-{2M\over r}\right){d^2\over dr^2} +{2(r-M)\over
r^2}{d\over dr} +\left({\omega^2\over\displaystyle
 1-{2M\over r}}-{\ell(\ell+1)\over r^2}\right)
\right]\phi^\nu(r)=0 \,. \label{eq:radial-eq2}
\end{eqnarray}
As long as we consider sufficiently large $\ell$, this radial wave
function can be completely specified up to an overall normalization
by the requirement that $\Phi^\nu:=(2z)^{-\nu}\phi_c^\nu$ does not
contain $\log z$ in its PN power series expansion with
respect to $z^2$ and $\epsilon/z$. The same condition
simultaneously determines the renormalized angular momentum
$\nu\approx \ell$. In fact, the equation for $\Phi^\nu$ becomes
\begin{eqnarray}
&&\left[z^2\partial_z^2+2z(\nu+1)\partial_z\right]\Phi^\nu
  =  \Biggl[{\epsilon\over z}
   (2-{\epsilon\over z})z^2 \partial_z^2+
    {\epsilon\over z} \left\{
    (4\nu+3)-{\epsilon\over z}(2\nu+1)\right\}z \partial_z\cr
 &&\qquad\qquad
+(\ell-\nu)(\ell+\nu+1)\left(1-{\epsilon\over z}\right)
       -z^2+{\epsilon\over z}\left(
           1-{\epsilon\over z}\right)\nu^2\Biggr]\Phi^\nu.
\end{eqnarray}
The right-hand side of this equation is of higher order in the PN
expansion. Substituting the Taylor expansion of $\Phi^\nu$ with
respect to $\epsilon/z$ and $z^2$ into the above equation, the
coefficients are determined order by order. However, the
$z$-independent terms in $\Phi^\nu$ vanish on the left-hand side.
Therefore, the terms which become zeroth order in $z$ should also
vanish on the right-hand side. This condition determines $\nu$
order by order. If, however, $\ell$ is not sufficiently large, terms
proportional to $z^{-2\ell-1}$ arise. For such terms, the left-hand
side is suppressed by a factor of $O(\epsilon^2)$. Hence, the
iteration scheme becomes invalid for small $\ell$, and we need to go
back to the original method presented in Ref.~\citen{ManoTakasugi}{}.

\section{A-Term}\label{app:A-term}

Here, we consider the $A$-term extracted from the $\tilde S$-part.
As we have explained in the main text, the $A$-term corresponds to
a jump of the field. Therefore, it is given by the antisymmetrized
Green function
\begin{eqnarray}
g_{\ell m\omega}^{\tilde S(-)}(r,r') = {-1 \over 2W_{\ell
m\omega}(\phi_c^\nu,\phi_c^{-\nu-1})} \left[ \phi_c^{\nu}(r)
\phi_c^{-\nu-1}(r') - \phi_c^{-\nu-1}(r) \phi_c^{\nu}(r') \right]
\,.
\end{eqnarray}
We introduce a new function, $\chi_c := r \phi_c$, which satisfies
\begin{eqnarray}
\left[\partial_{r^*}^2 + \omega^2 - V(r)\right] \chi_c (r^*) &=& 0
\,,
\label{psieq}\\
V(r) &=& \left(1-{2M \over r}\right) \left({\ell(\ell+1) \over
r^2} +{2(r-M)\over r^3} \right) \,,
\end{eqnarray}
where $r^* = r + 2M \ln(r/2M-1)$.
The Wronskian is written in
terms of $r^*$ and $\chi_c$ as
\begin{eqnarray}
W_{\ell m\omega}(\phi_c^\nu,\phi_c^{-\nu-1}) = \biggl({d\over
dr^*}\chi_c^{-\nu-1}(r^*)\biggr) \chi_c^{\nu}(r^*) -\biggl({d\over
dr^*}\chi_c^{\nu}(r^*)\biggr) \chi_c^{-\nu-1}(r^*) \,.
\end{eqnarray}

We expand the function $g_{\ell m\omega}^{\tilde S(-)}(r,r')$
in a power series with respect to $r^*-r^{*}{}'$ as
\begin{eqnarray}
g_{\ell m\omega}^{\tilde S(-)}(r,r')
&=& \sum_{n \geq 0} {\bm g}_n(r')(r^*-r^{*'})^n \,, \\
{\bm g}_n(r') &=& \left. {1 \over n!} \partial_{r^*}^n
  g_{\ell m\omega}^{\tilde S(-)}(r,r') \right|_{r=r'} \,.
\end{eqnarray}
The higher-order derivatives with respect to $\partial_{r^*}$ in
${\bm g}_n(r')$ can be reduced by using
Eq.~(\ref{psieq}). Hence, either one or zero $r^*$-derivatives remain
in the end.
Setting $r=r'$, the terms with no $r^*$
derivative vanish, while the terms with a single derivative yield
the Wronskian, which cancels the denominator. As a result,
we obtain a rather simple expression for ${\bm g}_n$.
 In fact, we have
\begin{eqnarray}
{\bm g}_1 &=& -{1 \over 2r'^2} \,, \cr {\bm g}_2 &=&
\left(1-{2M \over r'}\right) {1 \over 4r'^3} \,, \cr {\bm g}_3
&=& -{1 \over 12 r'^2} \left[ (-\omega^2 +V(r'))
-2\left(1-{2M\over r'}\right){r'-3M \over r'^3} \right] \,, \cr
 &\cdots& .
\nonumber
\end{eqnarray}

Note that only even positive integer powers of $\omega$ appear.
Therefore, let us consider terms proportional to $\omega^{2N}$ for a given $N$.
Because the factor $\omega^{2N}$ arises only from the elimination
of the $2N$ derivatives,
and because only a single derivative can remain at the end of the calculation,
$\omega^{2N}$ can be contained only in ${\bm g}_{n}(r')$ with
 $n\geq2N+1$.
Conversely, this means that we have 
\begin{eqnarray}
{\bm g}_{2N}(r')\,(r^*-r^*{}')^{2N}
&=& \sum_{n=0}^{N-1}\omega^{2n}\,a_n(r')(r^*-r^*{}')^{2N}\,,
\nonumber\\
{\bm g}_{2N+1}(r')\,(r^*-r^*{}')^{2N+1}
&=& \sum_{n=0}^{N}\omega^{2n}\,b_n(r')(r^*-r^*{}')^{2N+1}\,,
\end{eqnarray}
where $a_n(r')$ and $b_n(r')$ are independent of $\omega$.
 We replace $\omega$ with a time derivative, which acts on
$r'=z^r(t)$.
 The force is calculated by differentiating the potential once.
Hence, the term proportional to ${\bm g}_{2N}$ vanishes
in the coincidence limit, because it contains $2N-1$ derivatives at most.
As for the the term ${\bm g}_{2N+1}$, which contains $2N+1$
derivatives, each derivative
must act on each factor of $r^* -r^*{}'$ in $(r^*-r^*{}')^{2N+1}$
to give a finite result.
 Therefore, in ${\bm g}_{2N+1}(r')$, we
only need to keep $b_N$ in the coincidence limit. Thus,
we can simplify the coefficients ${\bm g}_n$ as
\begin{eqnarray}
   {\bm g}_{2N}\approx 0, \quad
    {\bm g}_{2N+1}\approx -{1\over 2 r'{}^2 (2N+1)!}(-\omega^2)^N.
\end{eqnarray}
Substituting this expression into Eq.~(\ref{forceS-S}), and summing
over the $m$-modes, the anti-symmetric part of the force can be
calculated as
\begin{eqnarray}
 F_{\alpha\ell}^{\tilde S(-)}
  & = & -\frac{q^2(2\ell+1)}{4\pi}\left[\partial_\alpha (r-z^r(t))
  \right]\sum_{n=0}^\infty
     {1\over 2{\cal E} z^r(t)^2}\left(1-\frac{2M}{z^r(t)}\right)^2
    \left({dz^r(t)/dt\over 1- 2M/z^r(t)} \right)^{2n}\cr
  & = &
     -{q^2{\cal L}{\cal E}
  \partial_\alpha (r-z^r(t)) \over 4\pi z^r(t) (z^r(t)-2M)
    (1+{\cal L}^2/z^r(t)^2)} ,
\end{eqnarray}
where we have used the equation of motion, Eq. (\ref{geodesic}).

\section{Absence of Large Powers of $\ell$}\label{app:large-ell}

We have $z=\omega r$ and $\epsilon=2M\omega,$ which contain
$\omega$ implicitly. This $\omega$ is replaced by a time
differentiation, which produces $m\Omega_\varphi$. That is,
$\omega$ is effectively of $O(\ell)$. In other words, we should
regard $z$ to be $O(\ell)$ and $\epsilon$ to be $O(\ell)$.
Then, considering the radial function given in
Eq.~(\ref{Phi}), one might think that the final expression for
the force would have terms with large positive powers of $\ell$.
 We explain here why this is actually {\bf NOT} the case, by analyzing
the leading $\ell$ behavior of $\Phi^\nu$ and $\Phi^{-\nu-1}$.

Let us introduce $\Psi^\nu$ by
\begin{equation}
 \Phi^\nu=\exp\left[\int^z dz \Psi^\nu\right]
=\exp\left[\int^r dr\,\omega\,\Psi^\nu\right].
\label{eq:phi0}
\end{equation}
Then the equation for $\Psi^\nu$ gives
\begin{equation}
\Psi^\nu=-\frac{\nu}{z}+\left[
\frac{\ell(\ell+1)}{z(z-\epsilon)}
-\frac{z^2}{(z-\epsilon)^2}
-\frac{\nu}{z(z-\epsilon)}
-\frac{2z-\epsilon}{z(z-\epsilon)}\Psi^\nu
-\partial_z\Psi^\nu\right]^{1/2}.
\end{equation}
Applying the large $\ell$ asymptotic expansion to this expression,
we have
\begin{equation}
\Psi^\nu=\Psi_0^\nu+\delta\Psi^\nu\,;
\quad
\Psi_0^\nu=
-\frac{\nu}{z}+\left[\frac{\ell^2}{z(z-\epsilon)}
-\frac{z^2}{(z-\epsilon)^2}\right]^{1/2},\label{eq:psi0}
\end{equation}
where $\Psi_0^\nu=O(\ell^0)$ and $\delta\Psi^\nu=O(1/\ell)$.

We find that, at the leading order in $\ell$, the condition that
the term of $O(1/z)$ in $\Psi_0^\nu$ should vanish determines
$\nu$ in the sense of the PN expansion:
\begin{eqnarray}
 \nu=\nu_0+O(\ell^0),\quad\nu_0=\ell-\frac{15\epsilon^2}{16\ell}+\cdots.
\end{eqnarray}
Also, $\Psi^\nu_0$ is given by
\begin{eqnarray}
 \Psi^\nu_0&=&-\frac{z}{2\ell}-\frac{z^3}{8\ell^3}-\frac{z^5}{16\ell^5}
+O(z^7/\ell^7)\cr &&
\,+\left(\frac{\ell^2}{2z^2}-\frac{3}{4}-\frac{5z^2}{16\ell^2}
+O(z^4/\ell^4)\right)\frac{\epsilon}{\ell}
\,+\left(\frac{3\ell^3}{8z^3}+O(z/\ell)\right)\frac{\epsilon^2}{\ell^2}
+\cdots.
\label{Psiexpand}
\end{eqnarray}
Substituting this expression into Eq.~(\ref{eq:phi0}), the result is
\begin{eqnarray}
 \Phi^\nu=1-\frac{z^2}{4\ell}-\frac{\ell\epsilon}{2z}+\frac{z^4}{32\ell^2}
+\frac{z\epsilon}{8}+\frac{\ell^2\epsilon^2}{8z^2}+\cdots,
\end{eqnarray}
which coincides with Eq.~(\ref{Phi}) in the large $\ell$ limit.
Another independent solution
$\Psi^{-\nu-1}_0$ can be obtained through the replacement
$\nu\to -\nu-1$, $\ell\to -\ell-1$, $+[\cdots]^{1/2}\to -[\cdots]^{1/2}$
in Eq.~(\ref{eq:psi0}).

The product of $\Phi^\nu(r)$ and
$\Phi^{-\nu-1}(r')$, which appears in the Green function, becomes
\begin{eqnarray}
&& \Phi^\nu (z)\Phi^{-\nu-1}(z')
=\exp\left(\int^z dz\,\Psi_0^\nu(z)+
           \int^{z'} dz'\,\Psi_0^{-\nu-1}(z')\right)
\times O(\ell^0)\,.
\end{eqnarray}
As clearly seen from the expanded form of $\Psi^\nu_0$
given in Eq.~(\ref{Psiexpand}),
we have
 $\Psi^\nu_0(z)=(\ell/\omega)\sum_{n=0}^\infty(\omega/\ell)^{2n}C_n(r)$,
 where the quantities $C_n$ are functions of $r$,
and $\Psi^{-\nu-1}_0(z)=-\Psi^\nu_0(z)+O(1/\ell)$.
This implies that, when $z$ and $z'$ are sufficiently close, we have 
\begin{eqnarray}
\int^z dz\,\Psi_0^\nu(z)+\int^{z'} dz'\,\Psi_0^{-\nu-1}(z')
=\ell\,(r-r')\sum_{n=0}^\infty(\omega/\ell)^{n}F_n(r,r')+O(\ell^0),\quad
\end{eqnarray}
where the $F_n$ are functions of $r$ and $r'$ that are
independent of $\omega$ and $\ell$ and regular in the
coincidence limit, $r\to r'$.
Therefore, we obtain
\begin{eqnarray}
 \Phi^\nu(z)\Phi^{-\nu-1}(z')=
\sum_{s=1}^\infty\sum_{n=0}^\infty
\ell^s(r-r')^s(\omega/\ell)^n \tilde F_{s,n}(r,r')+O(\ell^0),
\end{eqnarray}
where the $\tilde{F}_{s,n}$ are functions of $r$ and $r'$.
The terms apparently of $O(\ell^s)$ are always associated with
the factor $(r-r')^s$. Therefore, the
time differentiation must act on $r'=z^r(t')$ at least
$s$ times. Otherwise such terms vanish in the coincidence limit.
Each time differentiation, however, reduces the power
of $\omega$ by $1$, and hence it produces the factor $1/\ell$.
Thus, the terms that appear to be $O(\ell^s)$
turn out to be $O(\ell^0)$ in the end.


\begin{thebibliography}{99}

\bibitem{LIGO} LIGO web page: \texttt{http://www.ligo.caltech.edu/}

\bibitem{TAMA} TAMA300 web page: \texttt{http://tamago.mtk.nao.ac.jp/}

\bibitem{GEO} GEO600 web page: \texttt{http://www.geo600.uni-hannover.de/}

\bibitem{VIRGO}  VIRGO web page: \texttt{http://www.virgo.infn.it/}

\bibitem{LISA} LISA web page: \texttt{http://lisa.jpl.nasa.gov/}

\bibitem{Cutler}
C.~Cutler {\it et al.},
Phys.\ Rev.\ Lett.\
{\bf 70} (1993), 2984; astro-ph/9208005.

\bibitem{Blanchet:2002av}
L.~Blanchet,
Living Rev.\ Rel.\  {\bf 5} (2002), 3; gr-qc/0202016.

\bibitem{RW} T. Regge and J. A. Wheeler,
Phys. Rev. {\bf 108} (1957), 1063.

\bibitem{Zerilli}
F.~J.~Zerilli,
Phys.\ Rev.\ D {\bf 2} (1970), 2141.

\bibitem{Teukolsky}
S.~A.~Teukolsky,
Astrophys.\ J.\  {\bf 185} (1973), 635.

\bibitem{Chandra} S. Chandrasekhar, {\it Mathematical Theory of
Black Holes} (Oxford University Press, 1983).

\bibitem{supple}
Y.~Mino, M.~Sasaki, M.~Shibata, H.~Tagoshi and T.~Tanaka,
Prog.\ Theor.\ Phys.\ Suppl.\  No.{\bf 128} (1997), 1; gr-qc/9712057.

\bibitem{Sasaki-Tagoshi}
M.~Sasaki and H.~Tagoshi, Living Rev. Rel. (2003), to appear.;
 gr-qc/0306120.

\bibitem{DB} B. S. DeWitt and R.~W.~Brehme,
Annals Phys.\ {\bf 9} (1960), 220.

\bibitem{MST}
Y.~Mino, M.~Sasaki and T.~Tanaka,
Phys.\ Rev.\ D {\bf 55} (1997), 3457; gr-qc/9606018; 

Prog.\ Theor.\ Phys.\ Suppl.\  {\bf 128} (1997), 373; gr-qc/9712056.

\bibitem{QW}
T.~C.~Quinn and R.~M.~Wald,
Phys.\ Rev.\ D {\bf 56} (1997), 3381; gr-qc/9610053.

\bibitem{DW03}
S.~Detweiler and B.~F.~Whiting,
Phys.\ Rev.\ D {\bf 67} (2003), 024025; gr-qc/0202086.

\bibitem{ManoTakasugi} S.~Mano, H.~Suzuki and E.~Takasugi,
Prog.\ Theor.\ Phys.\ {\bf 95} (1996), 1079; gr-qc/9603020; 
Prog.\ Theor.\ Phys.\ {\bf 96} (1996), 549; gr-qc/9605057.

\bibitem{ManoTak}
S.~Mano and E.~Takasugi,
Prog.\ Theor.\ Phys.\ {\bf 97} (1997), 213; gr-qc/9611014.

\bibitem{Chrzanowski}
P.~L.~Chrzanowski,
Phys.\ Rev.\ D {\bf 11} (1975), 2042.

\bibitem{Ori:2002uv}
A.~Ori,
Phys.\ Rev.\ D {\bf 67} (2003), 124010; gr-qc/0207045.

\bibitem{MNS}
Y.~Mino, H.~Nakano and M.~Sasaki,
Prog.\ Theor.\ Phys.\  {\bf 108} (2003), 1039; gr-qc/0111074.

\bibitem{Barack:1999wf}
L.~Barack and A.~Ori,
Phys.\ Rev.\ D {\bf 61} (2000), 061502(R); gr-qc/9912010.

\bibitem{NMS}
H.~Nakano, Y.~Mino and M.~Sasaki,
Prog.\ Theor.\ Phys.\  {\bf 106} (2001), 339; gr-qc/0104012.

\bibitem{BMNOS}
L.~Barack, Y.~Mino, H.~Nakano, A.~Ori and M.~Sasaki,
Phys.\ Rev.\ Lett.\  {\bf 88} (2002), 091101; gr-qc/0111001.

\bibitem{Barack:2002mh}
L.~Barack and A.~Ori,
Phys.\ Rev.\ D {\bf 66} (2002), 084022; gr-qc/0204093.

\bibitem{Barack:2002bt}
L.~Barack and A.~Ori,
Phys.\ Rev.\ D {\bf 67} (2003), 024029; gr-qc/0209072.

\bibitem{Barack:2003mh}
L.~Barack and A.~Ori,
Phys.\ Rev.\ Lett.\  {\bf 90} (2003), 111101; gr-qc/0212103.

\bibitem{Detweiler:2002gi}
S.~Detweiler, E.~Messaritaki and B.~F.~Whiting,
Phys.\ Rev.\ D {\bf 67} (2003), 104016; gr-qc/0205079.

\bibitem{Burko:2000xx}
L.~M.~Burko,
Phys.\ Rev.\ Lett.\  {\bf 84} (2000), 4529; gr-qc/0003074.


\bibitem{Sago:2002fe}
N.~Sago, H.~Nakano and M.~Sasaki,
Phys.\ Rev.\ D {\bf 67} (2003), 104017; gr-qc/0208060.

\bibitem{NSS03}
H.~Nakano, N.~Sago and M.~Sasaki,; gr-qc/0308027.

\bibitem{Mino2003}
Y.~Mino,
Phys.\ Rev.\ D {\bf 67} (2003), 084027; gr-qc/0302075.

\bibitem{Jaranowski}
P.~Jaranowski and G.~Schafer,
Phys.\ Rev.\ D {\bf 57} (1998), 7274
[Errata {\bf 63} (2001), 029902]; gr-qc/9712075.

\bibitem{Gourgoulhon}
P.~Grandclement, E.~Gourgoulhon and S.~Bonazzola,
Phys.\ Rev.\ D {\bf 65} (2002), 044021; gr-qc/0106016.

\bibitem{Damour}
T.~Damour, B.~R.~Iyer and B.~S.~Sathyaprakash,
Phys.\ Rev.\ D {\bf 57} (1998), 885; gr-qc/9708034.

\end{thebibliography}
\end{document}